\documentclass[10pt, oneside, a4paper]{article}
\usepackage{latexsym}
\usepackage{amssymb}
\usepackage{amsthm}
\usepackage{cite}
\newcommand {\be} {\begin{equation}}
\newcommand {\ee} {\end{equation}}
\newcommand {\bea} {\begin{eqnarray}}
\newcommand {\eea} {\end{eqnarray}}
\newcommand {\bean} {\begin{eqnarray*}}
\newcommand {\eean} {\end{eqnarray*}}
\theoremstyle{plain}
\newtheorem{theorem}{Theorem}

\newcommand {\nn} {\nonumber}
\newcommand {\sss} {\scriptscriptstyle}
\newcommand {\scs} {\scriptstyle}

\newcommand {\we} {\wedge}
\newcommand {\fs}{\footnotesize}

\newcommand {\al} {\alpha}
\newcommand {\bet} {\beta}
\newcommand {\ga} {\gamma}
\newcommand {\Ga} {\Gamma}
\newcommand {\de} {\delta}
\newcommand {\De} {\Delta}
\newcommand {\ep} {\epsilon}

\newcommand {\om} {\omega}

\newcommand {\Si} {\Sigma}

\newcommand {\gh} {\ensuremath{gh_{\#}}}
\newcommand
{\funcright}[3]{\frac{\stackrel{\leftarrow}{\de}}{\de {#1^{\sss
#2}#3}}}
\newcommand
{\funcleft}[3]{\frac{\stackrel{\rightarrow}{\de}}{\de {#1_{\sss
#2}#3}}}

\newcommand{\rightd}[1]{\!\!\stackrel{\leftarrow}{\partial}_{\sss #1}}

\begin{document}
\begin{titlepage}
\vspace*{40mm}
\begin{center}
 {\LARGE\bf First order gauge field theories\\}
\end{center}\vspace*{0.0 mm}
\begin{center}
{\LARGE\bf from a superfield formulation}
\end{center}
\vspace*{3 mm}
\vspace*{3 mm}
\begin{center}Ludde Edgren\footnote{E-mail:
edgren@fy.chalmers.se} and Niclas Sandstr\"om\footnote{E-mail:
tfens@fy.chalmers.se}
 \\ \vspace*{7 mm} {\sl Department of Theoretical Physics\\ Chalmers University of Technology\\
G\"{o}teborg University\\
S-412 96  G\"{o}teborg, Sweden}\end{center} \vspace*{25 mm}
\begin{center}
\begin{abstract}
\vspace*{1 mm} Recently, Batalin and Marnelius proposed a
superfield algorithm for master actions in the BV-formulation for
a class of first order gauge field theories. Possible theories are
determined by a ghost number prescription and a simple local
master equation. We investigate consistent solutions of these
local master equations with emphasis on four and six dimensional
theories.
\end{abstract}
\end{center}
\end{titlepage}
\section{Introduction}
The Batalin-Vilkovisky (BV) formalism \cite{Batalin:1981ga, Batalin:1983qu} provides a powerful framework
for construction of quantum gauge field theories. In this
procedure the fundamental object is the so called master action,
consisting of fields and their antifields. A necessary consistency condition is
that this master action satisfies the master equation.
Recently it has been shown that many models allow for a simple form of
the master action in terms of superfields \cite{Abreu:1995su,
  Braga:1994su, Braga:1995su, Batalin:1997su,
  Batalin:1998su, Cattaneo:1999pa, Grigorev:1999su, Ikeda:2001de,
  Cattaneo:2001po, Alexandrov:1995ge, Cattaneo:2000hi,
  Cattaneo:2001on, Park:2000to, Batalin:2001ge, Batalin:2001su}. In
this paper we will follow the approach considered in
\cite{Batalin:2001ge, Batalin:2001su}. There Batalin and Marnelius
introduced a superfield algorithm for a class of first order gauge
field theories from a master action written in terms of
superfields. This type of master action was originally considered in \cite{Grigorev:1999su, Alexandrov:1995ge}. By means of a ghost number prescription this algorithm provides for a simple
and powerful method of finding all possible (local) terms entering
in such a master action. Furthermore, with appropriate boundary conditions this master
action can be expressed in terms of a local master action satisfying a much simpler local master equation.

This superfield algorithm leads to many consistent theories,
including the two dimensional Poisson sigma model \cite{Ikeda:1993tw, Schaller:1994po} and its superfield formulation considered by
Cattaneo and Felder \cite{Cattaneo:1999pa, Cattaneo:2001po},
Chern-Simons, topological Yang-Mills, BF-theories and
generalizations thereof. It also agrees with the general framework
set up by AKSZ \cite{Alexandrov:1995ge}.

In this communication we construct new consistent solutions of the
local master equation using (anti)canonical transformations. We
give explicit solutions to the master equation for first order
gauge field theories in four and six dimensions. In several papers
\cite{Ikeda:2001de, Barnich:1994lo, Barnich:1994re,
Boulanger:2000no, Barnich:2001se, Barnich:1993co, Henneaux:1997co,
Ikeda:2000de} it has been established that a large class of theories
are just trivial deformations of theories of a simpler kind. Our
results in six dimensions support these results. It seems that
only two interaction terms can account for a large class of first
order theories in six dimensions. Many of the other
possible terms (according to the superfield algorithm) belongs to
the same equivalence class and can thus be obtained by
(anti)canonical transformations from these two terms. We also
discuss some general feature of this class of gauge field theories
in various dimensions.

The paper is organized as follows. In the next section
we give a short review of the procedure proposed by Batalin and
Marnelius \cite{Batalin:2001ge, Batalin:2001su}. Section
\ref{solutions to} discusses the construction of solutions to the master
equation, for the class of first order gauge field theories considered in section
\ref{a specific}, using (anti)canonical transformations. In section \ref{sixdimensional} this
construction is applied to six dimensional theories. In
section \ref{gauge structure} we discuss how one can derive the
gauge transformations of the classical model that can be
extracted from a specific master action.
A discussion of some general features in various dimensions of the
class of quantum field theories studied in this paper is then given in
section \ref{structure of}. In the appendices we give notations,
conventions and some special results.

%
%
%
%
\section{Preliminaries}\label{a specific}
In \cite{Batalin:2001ge, Batalin:2001su} Batalin and Marnelius proposed a
superfield algorithm for the construction of master actions for a
class of first order gauge field theories. The master action  $\Si$ corresponding to
such an $n$-dimensional theory can be written as a field theory
living on a $2n$-dimensional supermanifold $\mathcal M$ where $n$
of the dimensions are Grassmann odd and $n$ Grassmann even,
 \be\label{masteraction}
    \Si[K^{\sss{P}}\!,K^*_{\sss{P}}]=\int_{\mathcal M}{d^{\sss n}ud^{\sss{n}}\tau
    \mathcal{L}_n(u,\tau)},
 \ee
where $K^{\sss P}$ and $K^*_{\sss P}$ are superfields. The supermanifold $\mathcal M$ is coordinatized by
$(u^a,\,\tau^a)$, where $a=\{1,...,n\}$ and $u^a$ denotes the Grassmann even and $\tau^a$
the Grassmann odd coordinates respectively. The Lagrangian density $\mathcal{L}_{\sss n}$ is of the form
 \be
    \mathcal{L}_n(u,\tau)=K^*_{\sss P}(u,\tau)DK^{\sss P}(u,\tau)(-1)^{\ep_{\sss{P}}\sss{+n}}-S(K^*_{\sss P}(u,\tau),K^{\sss
    P}(u,\tau)),
 \ee
where $D$ is the de Rham differential which locally can be written as \be\label{de_Rham}
    D:=\tau^a\frac{\partial}{\partial u^a}.
 \ee
It follows that $D$ is nilpotent
 \be
    D^2=0.
 \ee
Note that the terms in $S(K^*_{\sss P}(u,\tau),K^{\sss
P}(u,\tau))$ only contains superfields without derivatives. The
parities of the superfields $K^{\sss P}$ and associated superfields $K^*_{\sss P}$ are given by
 \bea
    \ep(K^{\sss P}) &:=& \ep_{\sss{P}}, \nonumber\\
    \ep(K^*_{\sss P}) &=& \ep_{\sss{P}}+1+n.
 \eea
Since the de Rham differential $D$ and $\tau^{\sss{a}}$ carries ghost
number one, the measure $d^{\sss{n}}\tau$ has ghost number $-n$. This
implies, since the master action $\Si$ has ghost number zero, that
 \be
   \gh K^{\sss P}+\gh K^*_{\sss P}=n-1.
 \ee
The local function $S$ possesses the following ghost number and
Grassmann grading
 \bea
    \gh(S) &=& n, \nonumber\\
    \ep(S) &=& n.
 \eea
The equations of motion that follows from the master action $\Si$
in (\ref{masteraction}) are
 \be
    DK^{\sss P}=(S,K^{\sss P})_{\sss n},\hspace{3mm} DK^*_{\sss P}=(S,K^*_{\sss P})_{\sss
    n}.
 \ee
They provide for a natural BRST-charge operator interpretation of the de Rham differential, defined in
 (\ref{de_Rham}). The local $n$-bracket $(\;\;,\;)_{\sss n}$
 introduced above, is defined by
 \be
 (F,G)_{\sss{n}}  =
 F\frac{\stackrel{\leftarrow}{\partial}}{\partial{K^{\sss{P}}}}\frac{\stackrel{\rightarrow}{\partial}}{\partial{K^*_{\sss{P}}}}G-(F\leftrightarrow
 G)(-1)^{\sss{(\ep(F)+n+1)(\ep(G)+n+1)}}.
 \ee
It satisfies the Jacobi identity
 \be\label{jacobin}
 ((F,G)_{\sss{n}},H)_{\sss{n}}(-1)^{\sss{(\ep(F)+n+1)(\ep(G)+n+1)}}+\mbox{cycle}(F,G,H)=0,
 \ee
 and has the graded symmetry property
 \be\label{gradesymm}
    (F,G)_{\sss{n}}=-(-1)^{\sss{(\ep(F)+n+1)(\ep(G)+n+1)}}(G,F)_{\sss{n}}.
 \ee
Due to (\ref{jacobin}) and (\ref{gradesymm}) we observe that
$(\,\,,\,)_{\sss{n}}$ is an "ordinary" antibracket in even
dimensions and a super Poisson bracket in odd dimensions. The
$n$-bracket carries $1$-$n$ units of ghost number
 \be\label{ghostcarry}
    \gh(F,G)_{\sss{n}}=\gh F+\gh G+1-n,
 \ee
and $n$+$1$ units of parity
 \be\label{paritycarry}
    \ep((F,G)_{\sss{n}})=\ep(F)+\ep(G)+n+1.
 \ee
It also satisfies the Leibniz rule
 \bea
    (F,GH)_{\sss{n}} & = &
    (F,G)_{\sss{n}}H+G(F,H)_{\sss{n}}(-1)^{\sss{\ep(G)(\ep(F)+n+1)}}, \nonumber\\
    (FG,H)_{\sss{n}} & = & F(G,H)_{\sss{n}}+(F,H)_{\sss{n}}G(-1)^{\sss{\ep(G)(\ep(H)+n+1)}}.
 \eea
Notice that the respective expansion of the superfields $K^{\sss P}$
and $K^*_{\sss P}$ in terms of the odd coordinates $\tau^{\sss{a}}$
lead to component fields which are either fields or antifields. Since the original fields
constitute the ghost number zero components of the superfields
$K^{\sss P}$ and $K^*_{\sss P}$, one obtains the following rules
for extracting the $n$ dimensional classical field theory
corresponding to a given master action $\Si$ of the form
(\ref{masteraction})
 \bea\label{limit}
    d^{\sss n}ud^{\sss n}\tau & \rightarrow & 1\nn\\
    D & \rightarrow & \mbox{exterior derivative } d\nn\\
    K^{\sss P}:\gh K^{\sss P}=k\geq 0 & \rightarrow & \mbox{k-form
    field}\,\,k^{\sss P} \mbox{where},\nn\\
    & &\ep(k^{\sss P})  = \ep_{\sss P}\mbox{\it\fs +k}\nn\\
    K^*_{\sss P}:\gh K^*_{\sss P}=(n-1-k)\geq 0 & \rightarrow &\mbox{\fs\it(n-1-k)}-\mbox{form field}\,\,
    k^*_{\sss P}\,\mbox{where},\nn\\
    & & \ep(k^*_{\sss P})=\ep_{\sss P}\mbox{\it\fs +k}\nn\\
    \mbox{all other superfields} & \rightarrow & 0\nn\\
    \mbox{pointwise multiplication} & \rightarrow & \mbox{wedge
    product}.
 \eea
%
%
\section{Solutions to the master equation}\label{solutions to}
In this section we consider the construction of theories having a
master action of the form \cite{Batalin:2001ge,Batalin:2001su},
 \bea
    \Sigma[K^*_{\sss{P}},K^{\sss{P}}]=\int\! d^{\sss{n}}ud^{\sss{n}}\tau
    K^*_{\sss{P}}(u,\tau)D
   K^{\sss{P}}(u,\tau)(-1)^{\ep_{\sss P}\sss{+n}}\!-S(K^*_{\sss{P}}(u,\tau),K^{\sss{P}}(u,\tau)).\nn
 \eea
The quantum master equation for a model $\Si$ is given by,
 \be
    \frac{1}{2}(\Si,\Si)=i\hbar\De\Si
 \ee
where the antibracket $(A,B)$ between two functionals is defined
by,
 \bea\label{globalbracket}
    (A,B)&:=&
    \int{A\funcright{K}{P}{(u,\tau)}(-1)^{\sss{(\ep_{\sss P}}\sss{n)}}d^{\sss{n}}u\;
    d^{\sss{n}}\tau\funcleft{K^*}{P}{(u,\tau)}B}\nn\\
    &&{}-(A\leftrightarrow
    B)(-1)^{\sss{(\ep(A)+1)(\ep(B)+1)}},
 \eea
and where the BV-Laplacian is given by,
 \be
    \De=\int{d^{\sss{n}}ud^{\sss{n}}\tau(-1)^{\sss{
    (n+1)\ep_{\sss{P}}}}\funcleft{K^{\sss
    P}}{}{(u,\tau)}\funcleft{K^*}{P}{(u,\tau)}}.
 \ee
In \cite{Batalin:2001su} it was shown that $\Si$ fulfills the
quantum master equation, provided that $S$ satisfies the local
$n$-{\it bracket master equation},
 \be \label{nbracket1}
    (S,S)_{\sss{n}}=0\label{n-bracketmaster},
 \ee
and the boundary condition
 \be \label{boundcond}
    \int{d^{\sss{n}}ud^{\sss{n}}\tau
    D\mathcal{L}(u,\tau)}=0\label{bound1}.
 \ee
Only when (\ref{nbracket1}) and (\ref{boundcond}) are satisfied
does $\Si$ represent a master action for a theory that
may be quantized in a consistent way. One may note that for
example Dirichlet boundary values on all the superfields trivially
satisfies (\ref{bound1}), but more general situations are also
allowed for.

Given a specific ansatz for the local master action $S$, the
problem is to find explicit expressions for the coefficients
of the superfields in $S$ that solves (\ref{n-bracketmaster}).
Finding such solutions by inserting a generic ansatz for $S$ into the
master equation are in general very difficult since it might lead
to equations for the coefficients in the ansatz that are too
complicated to solve. The general idea here is instead to start
with a local master action $S_0$ for which the solution is known,
and then perform canonical transformations\footnote{When mentioning
  canonical transformations in this paper, it is to be understood that
  we mean the transformations that preserves the local $n$-bracket,
  which are either canonical or anticanonical depending on $n$.} of $S_0$. If the local action $S_0$ is chosen
to have exactly the same superfield content as the original model
$S$, the canonically transformed $S_0$ will in several cases
constitute a solution for $S$. We will see examples of
this below.

A general (invertible) canonical transformation of an object $F$ is given by
 \be
    F_{\Ga}=e^{\mbox{\small ad}\,\Ga}F,
 \ee
where $\Ga$ is a canonical generator with adjoint action
 \be
    \mbox{ad}\,\Ga=(\,\cdot\,,\Ga)_{\sss{n}}.
 \ee

Instead of solving the local master equation (\ref{nbracket1}) directly
we propose now to use the canonical transformation
 \be\label{transaction}
    S_{\Ga}=e^{\ga\,\textrm{\small ad}\Ga}S_0=S_0+\ga{\left(S_0,\Gamma\right)}_{\sss{n}}+\frac{{\ga}^2}{2!}\left(\left(S_0,\Gamma\right)_{\sss{n}},\Gamma\right)_{\sss{n}}+\ldots,
 \ee
where $\ga$ denotes a real and even parameter. In terms of papers
\cite{Barnich:2001se, Barnich:1993co} $\ga$ is the deformation
parameter (in equation (\ref{transaction}) we have chosen to display
it explicitly instead of absorbing it into $\Ga$). Canonical
transformations correspond to trivial deformations - which implies
that they do not change the gauge structure. We have then
 \be
   (S_0,S_0)_{\sss{n}}=0 \quad \Rightarrow \quad (S_{\Ga},S_{\Ga})_{\sss{n}}=0.
 \ee
Since $\Ga$ is a canonical generator, it must preserve the ghost
number- and Grassmann gradings of the transformed quantity.  From
the properties (\ref{ghostcarry}) and (\ref{paritycarry}) of the
$n$-bracket, it follows then that we must impose the restrictions
 \bea\label{genproperties}
   gh_{\#}\Ga & = & n-1, \nonumber\\
   \ep(\Ga) & = & n+1 \,\,\,(mod 2).
 \eea
To be able to write down closed expressions for the
canonically transformed quantities we consider a subset of
canonical transformations, for which (\ref{transaction}) always
consists of a finite sum. To achieve this we consider the following
form of the generators that always will generate a canonical
transformation of the superfields which is first order in the parameter $\ga$ (see Appendix \ref{proof})
 \be\label{generator}
    \Ga=K^*_{\sss P_1}K^*_{\sss P_2}\ldots K^*_{\sss P_n}{\Ga^{\sss
    P_1P_2\ldots P_n}}_{\sss P_{1'} \sss P_{2'}\ldots P_{n'}}K^{\sss P_{1'}}K^{\sss P_{2'}}\ldots K^{\sss
    P_{n'}},
 \ee
where
 \be\label{whatever}
    \forall\,\, {\scriptstyle P}_i,{\scriptstyle P}_{i'}\,\,:\,{\scriptstyle P}_i \neq {\scriptstyle
    P}_{i'}.
 \ee
$K^{\sss P}$ and $K^*_{\sss P}$ are collective labels for
all superfields. The coefficients in (\ref{generator}), ${\Ga^{\sss
 P_1P_2\ldots P_n}}_{\sss P_{1'} \sss P_{2'}\ldots P_{n'}}$, may be
 functions of the ghost number zero superfields. Above $\Ga$ is chosen
  not to contain both a specific superfield and its associated
  superfield simultaneously. Note that the superfields in $\Ga$ must
  be chosen to satisfy the conditions in (\ref{genproperties}).

Note that since the master action $\Si_0$ in (\ref{masteraction}) transforms canonically as 
\bea
\Si_{{\bar{\Ga}}}&=&e^{{\textrm{\small ad\,}
    \bar{\Ga}}}\Si_0=\int d^{\sss{n}}ud^{\sss{n}}\tau\, e^{{\textrm{\small
      ad\,}\Ga}}\mathcal{L}_n\nn\\
&=&\int d^{\sss{n}}ud^{\sss{n}}\tau\, K^*_{\sss P}DK^{\sss P}(-1)^{\ep_{\sss{P}}\sss{+n}}-e^{{\textrm{\small
      ad\,}\Ga}}S_0+D\Ga,
\eea
it suffices to study canonical transformations of the local action $S_0$, provided
$ \int d^{\sss{n}}ud^{\sss{n}}\tau\,D\Ga=0$.
Above, $\bar{\Ga}=\int d^{\sss{n}}ud^{\sss{n}}\tau\,\Ga$ and $\Ga$ is
the local canonical generator defined by the properties in
(\ref{genproperties}, \ref{generator}) and (\ref{whatever}). Since
$\bar{\Ga}$ is a functional, $\textrm{\small ad\,}\bar{\Ga}$ is defined in terms of the functional antibracket (\ref{globalbracket}).

$S_0$ can often be written with only a few terms in such a way that it
trivially satisfies the local master equation. A suitable choice of $S_0$ will be
transformed to a $S_\Ga$ containing terms allowed by the general
ansatz for $S$, and as is shown in the examples given below, this
can be done for general forms of $S$. By comparing the
coefficients in $S$ and $S_\Ga$, the explicit solution to the
local master equation can be read off directly. In this way
solutions to very involved models in various dimensions can be
generated.

Before moving on to six dimensional gauge field theories in the next
section, let us illustrate the method for a class of models in four dimensions.
%
%
%
%
\subsection{Topological field theory in $n=4$} In a previous work by
Cattaneo and Felder \cite{Cattaneo:1999pa} the quantization of the
Poisson sigma model was studied. Their results were later
generalized by Batalin and Marnelius \cite{Batalin:2001ge} in the
superfield formulation of BV quantization considered here. When
they then generalized the method to any dimension in
\cite{Batalin:2001su} they considered as an application models in
$n=4$ that constitute a class of theories similar to Topological
Yang-Mills theories. Here we show that this theory is canonically
equivalent to a $2$-form self interacting type of theory in $n=4$.
The model considered in \cite{Batalin:2001su} is defined by the
local action
 \bea
    S& =&
    \frac{1}{2}T^*_{\sss E_1}T^*_{\sss E_2 }\om^{\sss E_1E_2}+\frac{1}{2}T^*_{\sss E_1}{\om^{\sss E_1}}_{\sss E_2E_3}T^{\sss E_2}T^{\sss E_3}+\nn\\
    &&{}+ \frac{1}{24}\om_{\sss E_1E_2E_3E_4}T^{\sss E_1}T^{\sss E_2}T^{\sss E_3}T^{\sss
    E_4},\label{n4CS}
 \eea
where the superfields carries the following ghost numbers and parities:
 \bea
    \gh(T^{\sss E}) &=& 1, \nonumber\\
    \gh(T^*_{\sss E}) &=&2, \nonumber\\
    \ep(T^{\sss E}) &=& 1, \nonumber\\
    \ep(T^*_{\sss E}) &=& 0.
 \eea
In order to find a solution to the local master equation for $S$,
i.e. $(S,S)_{\sss{n}}=0$, we consider the term,
 \be\label{turbo}
    S_0=T^*_{\sss E_1}T^*_{\sss E_2}\om^{\sss E_1E_2}.
 \ee
Obviously $S_0$ satisfies $(S_{\sss{0}},S_{\sss{0}})_{\sss 4}=0$.
The only possible\footnote{For sakes of simplicity we only consider
  canonical transformations that are non-vanishing in the limit (\ref{limit}).} canonical generator for $n=4$ which contains
only positive ghost number superfields is,
 \be\label{n4gen}
    \Ga=\frac{1}{3}\ga\,\ga_{\sss E_1E_2E_3}T^{\sss E_1}T^{\sss E_2}
    T^{\sss E_3}.
 \ee
Above $\ga$ is real and parametrizes the canonical transformation.
Due to the parity of the fields, and since
 \bea
   \ep(S_0) &=& 0, \nonumber\\
   \ep(\Ga) &=& 1,
 \eea
we have the following Grassmann gradings and symmetry properties
for the coefficients:
 \bea
    \ep(\ga_{\sss E_1E_2E_3}) &=& 0,\nonumber\\
    \ep(\om^{\sss E_1E_2})&=&0,\nonumber\\
    \om^{\sss E_1E_2} &=& \om^{\sss
    E_2E_1},\nonumber\\
    \ga_{\sss E_1E_2E_3} &=& -\ga_{\sss
    E_1E_3E_2}=\ga_{\sss
    E_3E_1E_2}.
 \eea
The canonically transformed action is given by
 \bea
    S_\Ga &=& T^*_{\sss E_1}T^*_{\sss E_2}\om^{\sss E_1E_2}-
    2\ga T^*_{\sss E_1}\om^{\sss E_1E}\ga_{\sss EE_2E_3}T^{\sss E_2}T^{\sss E_3}\nonumber\\
    &&{}+\ga^2\ga_{\sss E_1E_2E'}\om^{\sss E'E}\ga_{\sss
    EE_3E_4}T^{\sss E_1}T^{\sss E_2}T^{\sss E_3}T^{\sss E_4}.
 \eea
A comparison with (\ref{n4CS}) yields the following solution to the
master equation,
 \bea
    {\om^{\sss E_1}}_{\sss E_2E_3}&=&-4\ga\om^{\sss E_1E}\ga_{\sss
    EE_2E_3},\label{semi}\\
    \om_{\sss E_1E_2E_3E_4}&=&24\ga^2\ga_{\sss E_1E_2E}\om^{\sss
    EE'}\ga_{\sss E'E_3E_4}.\label{higherterm1}
 \eea
These solutions satisfy the master equation by construction - as
one may easily check by inserting them into to the equations
generated by $(S,S)_{\sss 4}=0$. The master equation for the
action (\ref{n4CS}) actually gives three equations - one of which
is enforcing the Jacobi-identities on the coefficients ${\om^{\sss
E_1}}_{\sss E_2E_3}$, which implies that they in general belong to
a super Lie algebra. From (\ref{semi}) we see that if $\om^{\sss
E_1E_2}$ is invertible, we may interpret it as a group metric.
Lowering the indices in equation (\ref{semi}) implies that ${\om_{\sss
E_1E_2E_3}}$ will be proportional to $\ga_{\sss E_1E_2E_3}$ and
thus totally antisymmetric with respect to all its indices - this
agrees with the fact that ${\om_{\sss E_1E_2E_3}}$ then belong to
a semi-simple Lie algebra. This is manifest in the solution
above.

Equations (\ref{semi},\ref{higherterm1}) display a generic feature
of solutions obtained in this way - all the generated coefficients
will be polynomials in terms of the coefficients used in the
canonical transformation and the coefficients in $S_0$. One should
also observe that this is also a solution to the more general case
when all the coefficients in equation (\ref{n4CS}) as well as
$\ga_{\sss E_1E_2E_3}$ are general functions of some ghost number
zero superfields. This will neither affect the master equation nor
the canonical transformations. The solution above obviously
degenerates when $\om^{\sss E_1E_2}=0$ but in this case the
coefficient will not be present in the original model (\ref{n4CS})
either. In such a situation one could use another $S_0$, for
example $S_0=\frac{1}{2}T^*_{\sss E_1}{\om^{\sss E_1}}_{\sss
E_2E_3}T^{\sss E_2}T^{\sss E_3}$, and proceed along the same lines
to find a solution. In the case of a non-invertible $\om^{\sss
  E_1E_2}$, a group metric interpretation is not possible. Hence,
$S_0$ in equation (\ref{turbo}) is only canonically equivalent 
to models whose coefficients can be factorized according to 
(\ref{semi}) and (\ref{higherterm1}). A careful analysis of each
given model (\ref{n4CS}) is required in order to decide whether this
is the case or not.

We observe that since the generator (\ref{n4gen}) is the only one
that contains exclusively superfields with positive ghost number
and that does not generate the identity transformation, it follows
trivially that the Lie-algebra term
 \be
    T^*_{\sss E_1}{\om^{\sss E_1}}_{\sss E_2E_3}T^{\sss E_2}T^{\sss E_3}
 \ee
is canonically inequivalent to the $2$-form self interacting type
of term,
 \be
    T^*_{\sss E_1}T^*_{\sss E_2}\om^{\sss E_1E_2}.
 \ee
Canonical inequivalence is however a necessary but $not$
sufficient condition for establishing the fact that two theories
have different gauge structure - for example, pure Maxwell theory
and Born-Infeld theory are canonically inequivalent, but do
possess the same gauge structure \cite{Barnich:2002sw}.
%
%
%
%
\section{$n=6$ Gauge field theories}\label{sixdimensional}
Consider the classical action for a general theory of
Schwarz type \cite{Schwarz:1978pa,Witten:1989qu} in $n=6$, i.e. a
nontrivial first order classical theory whose fields can be written entirely in terms of
differential forms. Such a general action consists of a
linear combination of terms with form degree 6, where the form
degree of the individual fields is spanning from 0 to 6. The
0-degree fields enters in the coefficients
of the other fields since the latter are functions. In the construction
below we will restrict ourselves to models which does not contain
any fields of form degree 6 and omit theories whose master action
have interaction terms\footnote{Interaction terms are those
contained in S, given the master action: $\Si[K,K^*]=\int{d^n
ud^n\tau (K^*DK-S[K,K^*])}$} containing fields with ghost number $5$. This
is a mild restriction and the reasons for this omission is
discussed at the end of this section.

Consider a classical action
 \be \label{classaction}
    S_{\sss Cl}[\Phi]=\int{d^{\sss{6}}x\mathcal{L}(\Phi(x))},
 \ee
which is a general functional of fields of form degree between 0 and 6. Now we
 will quantize this model in the superfield formalism according
to the algorithm described in \cite{Batalin:2001su}. The solution that
we will construct is a minimal one in the sense that the
corresponding master action $\Si$, which has the classical action
$S_{\sss Cl}$ as a limit, contains the smallest possible number of
fields. This we do for the sake of clarity. Extensions
are easy to construct and these will be discussed below.

Start by introducing $\{S^*_{\sss D},S^{\sss D},R^*_{\sss C},R^{\sss
  C},Q^*_{\sss B},Q^{\sss B}\}$ as the set of superfields, where the ghost number- and Grassmann ($\mathbb{Z}_2$) gradings of
the fields are given by
\begin{center}
\begin{tabular}{|c|c|c|c|c|c|c|}
  \hline
   & $S^{\sss D}$ & $S^*_{\sss D}$ & $R^{\sss C}$ & $R^*_{\sss C}$ & $Q^{\sss B}$ & $Q^*_{\sss B}$ \\
   \hline
  $\gh$ & $\fs 2$ & $\fs 3$ & $\fs 1$ & $\fs 4$ & $\fs 0$ & $\fs 5$ \\
  Parity & $\scs{\ep_{\sss{D}}}$ & $\scs{\ep_{\sss{D}}+1}$ & $\scs{\ep_{\sss{C}}}$ & ${\scs{{\ep_{\sss{C}}}}+1}$ & ${\scs{{\ep_{\sss{B}}}}}$ & ${\scs{{\ep_{\sss{B}}}}+1}$ \\ \hline
\end{tabular}
\end{center}
Observe that the fact $\gh Q^*_{\sss B}=5$ does not exclude $Q^*_B$ from
existing as a free field in the models we are studying. The most
general local action that can be written down under these conditions is,
 \bea\label{n6action}
  S&=&\om_{\sss{C_1} \sss{C_2} \sss{C_3} \sss{C_4} \sss{C_5}
  \sss{C_6}}R^{\sss{C_1}}R^{\sss{C_2}}R^{\sss{C_3}}R^{\sss{C_4}}R^{\sss{C_5}}R^{\sss{C_6}}\nn\\
  &&{}+\om_{\sss{C_1} \sss{C_2} \sss{C_3} \sss{C_4}
  \sss{D_1}}R^{\sss{C_1}}R^{\sss{C_2}}R^{\sss{C_3}}R^{\sss{C_4}}S^{\sss{D_1}}\nn\\
  &&{}+\om_{\sss{D_1} \sss{D_2} \sss{C_1} \sss{C_2}}S^{\sss{D_1}}S^{\sss{D_2}}R^{\sss{C_1}}R^{\sss{C_2}}+\om_{\sss{D_1} \sss{D_2}
  \sss{D_3}}S^{\sss{D_1}}S^{\sss{D_2}}S^{\sss{D_3}}\nn\\
  &&{}+S^*_{\sss{D_1}}S^*_{\sss{D_2}}\om^{\sss{D_1}
    \sss{D_2}}+R^*_{\sss{C_1}}{\om^{\sss{C_1}}}_{\sss{D_1}}S^{\sss{D_1}}+R^*_{\sss{C_1}}{\om^{\sss{C_1}}}_{\sss{C_2} \sss{C_3}}R^{\sss{C_2}}R^{\sss{C_3}}\nn\\
  &&{}+S^*_{\sss{D_1}}{\om^{\sss{D_1}}}_{\sss{C_1} \sss{C_2}
  \sss{C_3}}R^{\sss{C_1}}R^{\sss{C_2}}R^{\sss{C_3}}+S^*_{\sss{D_1}}{\om^{\sss{D_1}}}_{\sss{C_1}
  \sss{D_2}}R^{\sss{C_1}}S^{\sss{D_2}}.
 \eea
All the various coefficients $\om$ above are functions of the
ghost number zero superfield $Q^{\sss B}$. The parities and symmetry
properties of the coefficients will not be given here since they
follow directly from the parity of the superfields and the fact that $\ep(S)=0$. Starting with the action
 \be\label{n6s0}
    S_0=R^*_{\sss{C_1}}{\omega^{\sss{{C}_1}}}_{\sss{D_1}}S^{\sss{D_1}}+S^*_{\sss{D_1}}S^*_{\sss{D_2}}\omega^{\sss{D_1}
    \sss{D_2}},
 \ee
all possible terms in (\ref{n6action}) can be generated using
canonical transformations. The two terms in the action
(\ref{n6s0}) is canonically inequivalent as can easily be seen by
writing down all possible canonical generators with positive ghost
number (there are only four of them). The requirement that $S_0$
satisfies the master equation implies,
 \be\label{requirement}
    {\omega^{\sss{{C}_1}}}_{\sss{D}}\omega^{\sss{D} \sss{D_2}}=0.
 \ee
This means that ${\omega^{\sss{C_1}}}_{\sss{D_1}}$ can be
constructed from the null vectors of $\omega^{\sss{D_1} \sss{D_2}}$
or vice versa. Furthermore, we observe that the invertibility of
one of these two coefficients implies the vanishing of the other.
Thus, there exist no solution to the master equation for the
action (\ref{n6action}) where both
${\omega^{\sss{C_1}}}_{\sss{D_1}}$ and $\omega^{\sss{D_1}
\sss{D_2}}$ are invertible, since equation (\ref{requirement})
will be one of the equations imposed by $(S,S)_6=0$. The
superfield content of $S$ can be obtained from $S_0$ by using the
following canonical generator
 \be
    \Gamma=\alpha\,\Gamma_1+\beta\,\Gamma_2+\gamma\,\Gamma_3,
 \ee
where
  \bea
    \Gamma_1&=&S^*_{\sss{D_1}}{\ga^{\sss{D_1}}}_{\sss{\sss{C}_1} \sss{{C}_2}}R^{\sss C_1}R^{\sss C_2},\nonumber\\
    \Gamma_2&=&\ga_{\sss{{C}_1} \sss{C_2} \sss{C_3} \sss{C_4} \sss{C_5}}R^{\sss{C_1}}R^{\sss{C_2}}R^{\sss{C_3}}R^{\sss{C_4}}R^{\sss{C_5}},\nonumber\\
    \Gamma_3&=&\ga_{\sss{C_1} \sss{D_1}
    \sss{D_2}}R^{\sss{C_1}}S^{\sss{D_1}}S^{\sss{D_2}}.
 \eea
Although $\Ga_1, \Ga_2$ and $\Ga_3$ are not the only possible
choice in $n=6$ using only positive ghost number superfields - they
are sufficient. Note that all the coefficients
${\ga^{\sss{D_1}}}_{\sss{\sss{C}_1C_2}}$, $\ga_{\sss{{C}_1} \sss{C_2}
\sss{C_3} \sss{C_4} \sss{C_5}}$, $\ga_{\sss{C_1} \sss{D_1D_2}}$
and the even and real parameters $\al, \bet$, $\ga$ are functions of the ghost
number zero superfield $Q^{\sss B}$. An identification of the solution to the master equation $(S,S)_6=0$ can be done by comparing the canonically
transformed action $S_{\Gamma}$ with $S$. We find,
   \bea\label{solution}
 {\om^{\sss{C_1}}}_{\sss{C_2} \sss{C_3}}&=&\al {\om^{\sss{C_1}}}_{\sss{D}}{\ga^{\sss{D}}}_{\sss{C_2} \sss{C_3}}\nn,\\
 {\om^{\sss{D_1}}}_{\sss{C_1} \sss{C_2} \sss{C_3}}&=&-2\al^2{\ga^{\sss{D_1}}}_{\sss{C_1} \sss{C}}{\om^{\sss{C}}}_{\sss{D}}{\ga^{\sss{D}}}_{\sss{C_2}
   \sss{C_3}}\nn,\\
 {\om^{\sss{D_1}}}_{\sss{C_1} \sss{D_2}}&=&-2\al{\ga^{\sss{D_1}}}_{\sss{C_1}
   \sss{C}}{\om^{\sss{C}}}_{\sss{D_2}}\nn\\
 &&{}-4\ga\om^{\sss{D_1} \sss{D}}\ga{\sss{D} \sss{D_2}
   \sss{C_1}}\nn,\\
 \om_{\sss{D_1} \sss{D_2} \sss{D_3}}&=& -\ga \ga{\sss{D_1} \sss{D_2}
   C}{\om^{\sss{C}}}_{\sss{D_3}}\nn,\\
 \om_{\sss{D_1} \sss{D_2} \sss{C_1}\sss{C_2}}&=&{}-\ga\al\ga{\sss{D_1} \sss{D_2}
  \sss{C}}{\om^{\sss{C}}}_{\sss{D}}{\ga^{\sss{D}}}_{\sss{C_1}\sss{C_2}}\nn\\
&&{}-4\ga^2{\ga}_{\sss{C_1} \sss{D_1 D}}\om^{\sss{D
D'}}{\ga}_{\sss{D'
  C_2}\sss{D_2}}\nn\\
  &&{}+4\al\ga\ga_{\sss{C_1} \sss{D_1 D}}{\ga^{\sss{D}}}_{\sss{C_2
  C}}{\om^{\sss{C}}}_{\sss{D_2}}\nn,\\
  \om_{\sss{C_1} \sss{C_2} \sss{C_3}\sss{C_4}\sss{C_5} \sss{C_6}}&=&-5\al\bet\ga_{\sss{C_1} \sss{C_2} \sss{C_3} \sss{C_4}
    \sss{C}}{\om^{\sss{C}}}_{\sss{D}}{\ga^{\sss{D}}}_{\sss{C_5} \sss{C_6}},\nn\\
  \om_{\sss{C_1} \sss{C_2} \sss{C_3} \sss{C_4}
  \sss{D_1}}&=&-5\bet\ga_{\sss{C_1} \sss{C_2} \sss{C_3} \sss{C_4
  C}}{\om^{\sss{C}}}_{\sss{D_1}}\nn\\
  &&{}+4\ga\al^2\ga_{\sss{C_1} \sss{D_1 D}}{\ga^{\sss{D}}}_{\sss{C_2
    C}}{\om^{\sss{C}}}_{\sss{D'}}{\ga^{\sss{D'}}}_{\sss{C_3}
    \sss{C_4}}.
 \eea
The solution given here is for the sake of clarity given for the
case when all the superfields $\{S^{\sss D},R^{\sss C},Q^{\sss
B}\}$ have even parity. For the complete solution with arbitrary
parities on the superfields we refer to Appendix \ref{container}.
Given a classical model fulfilling the restrictions given above,
a quantum master action having (\ref{classaction}) as its
limit is defined by (\ref{n6action}) and the solution to the
coefficients (\ref{solution}). One can extend this model by
introducing higher order ghost number superfields into the action,
and for original theories possessing a high degree of reducibility
this is indeed necessary in order to construct the quantum theory.
The reason we omitted fields of form degree $6$ was that it will
introduce superfields with negative ghost number into the theory, due
to the relation
 \be
    \gh K^{\sss P}+\gh K^*_{\sss P}=n-1.
 \ee
In that case the master action and the canonical generators are
allowed to contain an arbitrary number of superfields and for
brevity we choose here to solve for a model excluding these
superfields. Extensions to ghost number $6$ superfields and higher
does not bring in any extra complications with respect to how the
solutions are constructed. All these models will have the same
classical limit since all the terms containing negative ghost
number superfields will be set to zero according to (\ref{limit}).
The reason for also excluding the superfields $Q^{\sss B}$, with
ghost number five, is that they couple to all the coefficients
$\om$ in the master equation, since they are functions of the
scalar superfields $Q^{\sss B}$. This implies in general that one
must solve complicated equations for the coefficients in $S_0$.
Finding a way to construct solutions in that case would be a
natural generalization of the method described above.

Note that the solution (\ref{solution}) is derived under the
assumption of non-vanishing coefficients in (\ref{n6action}). For
example, a local action with only the ${\omega^{{\sss C_1}}}_{\sss{D_1D_2}}$ term nonzero is obviously not canonically equivalent to the local
action (\ref{n6s0}). The question of which actions of the
form (\ref{n6action}) that are canonically equivalent to simpler actions
like (\ref{n6s0}), is in general very difficult to answer. One
rigorous but difficult way to investigate such a classification would
be to study the cohomology of the BRST-operator $D=(S\,,\,.)$ along the lines
of Henneaux et al. in for example \cite{Barnich:1993co, Barnich:1993in, Henneaux:1997al}

 At the classical level, for every given model, the possible number
of interaction terms involving fields of higher form
degree\footnote{Higher meaning of degree $n-1$ or $n$ where $n$ is
the dimension of the manifold on which the original model is
formulated.} are very few - and this for obvious reasons. This
fact explains in part the reason why some of the most interesting
models studied in the literature, such as Yang-Mills, BF-theories
and Chern-Simons, exhibit absence of fields of higher form degree.
%
%
%
%
%
%
%
%
%
%
%
%
%
%
%
%
%
%
\section{Gauge structure of the master action}\label{gauge
structure} In this section we investigate the connection between
the master action and the corresponding classical action. In
\cite{Batalin:2001su} it was shown that the $\Si$ variations of
the superfields are given by,
 \bea\label{derivegauge}
    &&{}\de_{\Si}K^{\sss P}{}={}(\Si,K^P)={}(-1)^{\sss n}(DK^{\sss
    P}-(S,K^{\sss P})_{\sss n}),\nn\\
    &&{}\de_{\Si}K^*_{\sss P}{}={}(\Si,K^*_{\sss P}){}={}(-1)^{\sss n}(DK^*_{\sss
    P}-(S,K^*_{\sss P})_{\sss n}).
 \eea
These variations can be used to determine the gauge
transformations of the classical model $S_{\sss Cl}$ corresponding
to the master action $\Si$. This is done by applying the rules
(\ref{limit}) to the $\Si$-variations (\ref{derivegauge}) and then
replacing {\it each} $k$-form field ($K^P$ reduces to a k-form field) in every term by a
($k$-$1$)-form gauge parameter, one at a time. In the case of
$0$-form fields (scalars) the corresponding gauge parameter is
zero. For example, from the following $\Si$-variation of a field
$S^*_{\sss D}$ in $n=4$,
 \bea
    \de_{\Si}S^*_{\sss D}{}={}(\Si,S^*_{\sss D}){}={}DS^*_{\sss
    D}-T^*_{\sss E_1}{\om^{\sss E_1}}_{\sss E_2E_3D_1D}T^{\sss E_2}T^{\sss E_3}S^{\sss
    D_1},
 \eea
where $S^*,T^*,T,S$ are ghost number $3,2,1$ and $0$ - fields
respectively, we derive the gauge transformation,
 \bea\label{classgauge}
    \de s^*_{\sss D} &=& d\tilde{s}^*_{\sss D}-
    \tilde{t}^*_{\sss E_1}{\om^{\sss E_1}}_{\sss E_2E_3D_1D}t^{\sss E_2}t^{\sss E_3}s^{\sss
    D_1}\nn\\
    &&{}- 2t^*_{\sss E_1}{\om^{\sss E_1}}_{\sss E_2E_3D_1D}\tilde{t}^{\sss E_2}t^{\sss E_3}s^{\sss
    D_1},
 \eea
of the corresponding classical $3$-form field, $s^*_{\sss{D}}$. Above, the tilde
quantities denotes the $(k$-$1)$-form gauge parameters
corresponding to the $k$-form fields without tilde. Note, that there
is no gauge parameter corresponding to the scalar field $s^{\sss
D}$ in (\ref{classgauge}). In section \ref{klasa} below, explicit
examples of gauge transformations derived from (\ref{derivegauge})
is given.

Now, one might ask the question: given a classical action with
certain interaction terms - what class of master actions reduce to
this classical action in the limit (\ref{limit})? It turns
out that this class is defined by a specific choice of the
parities of the superfields entering the master action. These
constraints originates from the symmetries of terms in the master
action which are of order quadratic or higher in any specific
superfield - which implies, since we always can commute these
fields in the graded sense, that the coefficients must possess a
(graded) symmetry in order to be non vanishing. It is obvious that
it suffices to look at quadratic terms in order to derive the
parity constraints. Let us look at the terms containing associated
superfields $K^*$ first - for example a term
 \be
    K^*_{\sss P_1}K^*_{\sss P_2}\om^{\sss P_1P_2},
 \ee
implies the following symmetry of the coefficient,
 \be\label{symm1}
    \om^{\sss P_1P_2}=(-1)^{\sss ({\scriptstyle \ep}_{P_1}+1+n)({\scriptstyle \ep}_{P_2}+1+n)}\om^{\sss P_2P_1}.
 \ee
This term reduces to
 \be
    k^*_{\sss P_1}\wedge k^*_{\sss P_2}\om^{\sss P_1P_2},
 \ee
where according to form degree- and parity description of
(\ref{limit})
 \be\label{antirule}
    k^*_{\sss P_1}\wedge k^*_{\sss P_2}=(-1)^{\sss
    ({\scriptstyle \ep}_{P_1}+k)({\scriptstyle \ep}_{P_2}+k)+n+1+k} k^*_{\sss P_2}\wedge k^*_{\sss P_1}.
 \ee
Above, $k$ denotes the form degree of the corresponding field
$k^{\sss P}$. In deriving (\ref{antirule}) we used the following
property of two $\mathbb{Z}_2$-graded $r$- and $s$-forms $A$ and
$B$ with parities $a$ and $b$ respectively,
 \be
    A\wedge B=B\wedge A(-1)^{\sss ab+rs}.
 \ee
Thus if the coefficient $\om^{\sss P_1P_2}$ is to be non-vanishing
in the classical action for both the symmetries defined by
(\ref{symm1}) and (\ref{antirule}), we must have
  \be\label{caseevenanti}
    {\scriptstyle (\ep_{P_1}+\ep_{P_2})(n+1)=k(\ep_{P_1}+\ep_{P_2})}.
 \ee
A completely analogous analysis of the field term
 \be
  \om_{\sss P_1P_2}K^{\sss P_1}K^{\sss P_2},
 \ee
gives the single condition
 \be\label{casefield}
    {\scriptstyle k(\ep_{P_1}+\ep_{P_2})=0}.
 \ee
We conclude that the only way to satisfy all cases
(\ref{caseevenanti}) and (\ref{casefield}) in all dimensions $n$,
is to require that all fields $K^{\sss P}$ of a specific $\gh$
have the same parity,
 \be
    \forall{\scriptstyle P_1,P_2\,:\,\ep_{P_1}+\ep_{P_2}=0}.
 \ee
This requirement ensure the existence of a classical term having
the same symmetry as its corresponding term in the master action.
It should be noted however, that some terms may still vanish at
the classical level due to the algebra structure of the model in
question. This is the case for certain terms with Lie-algebra
valued fields that is traced over, such as for example $n$-th
power terms of $1$-form fields in even dimensions $n$.
\subsection{Classical theory} \label{klasa}
We will now take a look at how one can derive the gauge
transformations for a classical model, from its corresponding
master action. Consider the six dimensional model discussed in
section \ref{sixdimensional} and whose interaction terms are given
in (\ref{n6s0}). The master action of this theory is given by
 \bea
 \Si&=&\int_{\mathcal M}{d^{\sss
 n}}ud^{\sss{n}}\tau\,\Big\{S^*_{\sss D}DS^{\sss {D}}(-1)^{\sss \ep_D}+R^*_{\sss C}DR^{\sss C}(-1)^{\sss \ep_C}+Q^*_{\sss B}DQ^{\sss
 B}(-1)^{\sss \ep_B}\nn\\
 &&\hspace{20mm}-(R^*_{\sss C}{\om^{\sss C}}_{\sss D}S^{\sss D}
 +S^*_{\sss D_1}S^*_{\sss D_2}\om^{\sss D_1D_2})\Big\}.
 \eea
Note that the coefficients in general depend on the scalar fields
$Q^{\sss B}$,
 \bea
    {\om^{\sss C}}_{\sss D} &=& {\om(Q)^{\sss C}}_{\sss D},\\
    \om^{\sss D_1D_2} &=& \om(Q)^{\sss D_1D_2}.
 \eea
By performing the reduction (\ref{limit}) we obtain the following
classical action,
 \bea
    \Si_{Cl}&=&\int{\Big\{s^*_{\sss D}\we ds^{\sss D}(-1)^{\sss \ep_D}+r^*_{\sss C}\we dr^{\sss C}(-1)^{\sss \ep_C}+q^*_{\sss B}\we dq^{\sss
    B}(-1)^{\sss \ep_B}}\nn\\
    &&\hspace{7mm}-(r^*_{\sss C}\we{\om^{\sss C}}_{\sss D}\we s^{\sss D}
    +s^*_{\sss D_1}\we s^*_{\sss D_2}\om^{\sss D_1D_2})\Big\}.
 \eea
Above, $q,r,s,s^*,r^*,q^*$ are $0,1,2,3,4$ and $5$-form fields
respectively. The gauge invariance of the classical model is
easily derived from (\ref{derivegauge}) and is given by,
 \bea
    \de s^*_{\sss D}&=&d\tilde{s}^*_{\sss D}-\tilde{r}^*_{\sss
    C}{\om^{\sss C}}_{\sss D},\\
    \de s^{\sss D}&=&d\tilde{s}^{\sss D}+2(-1)^{\sss \ep_D+1}\tilde{s}^*_{\sss D_1}\om^{\sss
    DD_1},\\
    \de r^*_{\sss C}&=&d\tilde{r}^*_{\sss C},\\
    \de r^{\sss C}&=&d\tilde{r}^{\sss C}+(-1)^{\sss \ep_C+1}{\om^{\sss
    C}}_{\sss D_1}\tilde{s}^{\sss D_1},\\
    \de q^*_{\sss B}&=&d\tilde{q}^*_{\sss B}+\tilde{r}^*_{\sss C}{\om^{\sss C}}_{\sss
    D}\rightd{B}s^{\sss D}(-1)^{\sss \ep_D\ep_B}+r^*_{\sss C}{\om^{\sss
    C}}_{\sss D}\rightd{B}\tilde{s}^{\sss D}(-1)^{\sss
    \ep_D\ep_B}\nn\\
    &&+2\tilde{s}^*_{\sss D_1}{s}^*_{\sss D_2}\om^{\sss
    D_1D_2}\rightd{B},\\
    \de q^{\sss B}&=&0.
 \eea
The tilded quantities $\tilde{r}, \tilde{s}, \tilde{s}^*, \tilde{q}^*$
and $\tilde{r}^*$ denotes $0,1,2,3$ and $4$-form gauge
parameters respectively. The gauge structure above can be viewed
as a consistent deformation of the gauge structure of the abelian
BF theory \cite{Ikeda:2001de},
 \bea
    \Si_{Cl}&=&\int s^*_{\sss D}\we ds^{\sss D}(-1)^{\sss \ep_D}+r^*_{\sss C}\we dr^{\sss C}(-1)^{\sss \ep_C}+q^*_{\sss B}\we dq^{\sss
    B}(-1)^{\sss \ep_B},\nn
 \eea
possessing the gauge symmetry
 \bea\label{abeliangauge}
    \de s^*_{\sss D}=d\tilde{s}^*_{\sss D}, & \de s^{\sss D}=d\tilde{s}^{\sss D},\\
    \de r^*_{\sss C}=d\tilde{r}^*_{\sss C}, & \de r^{\sss C}=d\tilde{r}^{\sss C},\\
    \de q^*_{\sss B}=d\tilde{q}^*_{\sss B}, & \de q^{\sss B}=d\tilde{q}^{\sss B}.
 \eea
In fact, from the relatively recent accomplishments of Ikeda
\cite{Ikeda:2001de,Ikeda:2000de} it is clear from the rules
(\ref{limit}) above that {\it all} classical models derived from a
master action of the form (\ref{masteraction}) represent
consistent deformations of abelian BF-theories.
\section{Structure of the master action in various
dimensions}\label{structure of} There are certain general rules
and similarities existing in various dimensions for the theories
we are studying. The first simple analysis one can do, is to look
for what constraints the requirement $gh_{\#}S=n$ impose on the
coefficients in the local master action $S$. A Yang-Mill's term in
the action always has one of the following two structures,
 \be
    S_{YM}=K^*_{\sss{P_1}}{\om^{\sss{P_1}}}_{\sss{P_2P_3}}K^{\sss{P_2}}K^{\sss{P_3}}\label{lie1},
 \ee
or
 \be
    S_{YM}=K^*_{\sss{P_1}}K^*_{\sss{P_2}}{\om^{\sss{P_1P_2}}}_{\sss{P_3}}K^{\sss{P_3}}\label{lie2},
 \ee
depending in what dimension the theory is formulated. The master
equation will impose Jacobi-identities for the coefficients
${\om^{\sss{P_1}}}_{\sss{P_2P_3}}\,
({\om^{\sss{P_1P_2}}}_{\sss{P_3}})$, thus demanding them to belong
to a Lie superalgebra. We observe from the superfield
ghost number relation
 \be
   \gh K^{\sss{P}}+\gh K^*_{\sss{P}}=n-1,
 \ee
that one of the superfields in $S_{YM}$ always have ghost number
one, implying that we always can identify the corresponding field
with a $1$-form gauge field. It follows that terms with the
structure $S_{YM}$ are allowed in all dimensions. The existence of
a group metric requires terms of one of the following two forms in
the master action,
 \bea \label{group1}
    S_{GM} & = &\om_{\sss{P_1P_2}}K^{\sss{P_1}}K^{\sss{P_2}},
 \eea
or
 \bea \label{group2}
   S_{GM} & = & K^*_{\sss{P_1}}K^*_{\sss{P_2}}\om^{\sss{P_1P_2}}.
 \eea
This fact follows from the requirement that it must be a quadratic
term with $\gh=n$ which couples, via the master equation, to any
of the Lie superalgebra terms (\ref{lie1},\ref{lie2}). A term with
one $K$- and one $K^*$-superfield is ruled out due to the $\gh(S)=n$
condition.

The ghost number requirement for the terms
(\ref{group1},\ref{group2}) together with a specific term
$S_{YM}$, yields $n=2$ or $n=4$. Thus, a quadratic coefficient in
the action can only have a group metric interpretation in the
first two even dimensions. In these two cases, an invertible
$\om_{\sss{P_1P_2}}$ or $\om^{\sss{P_1P_2}}$ can be used to raise
and lower the indices of the Lie superalgebra coefficients
(\ref{lie1},\ref{lie2}).

The number of possible (positive ghost number) interaction terms
that can exist in a master action $\Si$ in dimension $n$, is given
by the function $T(n)$,
 \[
    T(n)=\left\{
        \begin{array}
            {l@{\quad\quad}l}
            P(n) & n \mbox{ even}\\
            P(n)+P([\frac{n}{2}]+1)+\de_{n3}-\de_{n1} & n \mbox{ odd.}
        \end{array}
        \right.
 \]
Above $P(n)$ denotes the $Partition function$ for the integer $n$
and $[\frac{n}{2}]$, the integer part of $n/2$ . For the first
$11$ integers $T(n)$ and $P(n)$ is given by,
\begin{center}
\begin{tabular}{|c|c|c|c|c|c|c|c|c|c|c|c|}\hline
    $n$ & 1 & 2 & 3 & 4 & 5 & 6 & 7 & 8 & 9 & 10 & 11 \\\hline
 $T(n)$ & 1 & 2 & 6 & 5 & 10 & 11 & 20 & 22 & 37 & 42 & 67  \\ \hline
 $P(n)$ & 1 & 2 & 3 & 5 & 7 & 11 & 15 & 22 & 30 & 42 & 56  \\ \hline
\end{tabular}
\end{center}
The relatively large number of possible terms in the master
actions in odd dimensions, stems from the fact that it always
exist superfields $K$ and $K^*$ with the same ghost number when
$n$ is odd, thus leading to more possible combinations of
interaction terms. If negative ghost number superfields are
included, the number of possible terms is of course infinite.
\section{Acknowledgments}
We would like to thank Robert Marnelius for valuable comments.
\begin{appendix}
%
%
%
%
\section{Notation}
 \bean
 n &=& \mbox{dimension of the base manifold of the original theory,}\\
 K &=& \mbox{collective label of superfields,}\\
 K^* & = & \mbox{collective label of associated superfields to K,}\\
 K^{\sss{P}} &=& \mbox{arbitrary superfield,}\\
 K^*_{\sss{P}} &=& \mbox{arbitrary associated superfield,}\\
 \ep(K^{\sss{P}}) &:=& \ep_{\sss{P}}\,\,\mbox{= Grassmann parity of superfield}\,K^{\sss{P}},\\
 \ep(K^*_{\sss{P}}) &=& \ep_{\sss{P}}+n+1\,\,\mbox{= Grassmann parity of
 associated superfield}\,K^*_{\sss{P}},\\
 \gh &=& \mbox{ghost number,}\\
 \ep(F) &=& \mbox{Grassmann parity of object F}.\\
 \eean
%
%
%
%
\section{Conventions}
Ghost numbers of the superfields are chosen according to the convention
 \bea
    \gh K^*_{\sss{P}}  \geq  \gh K^{\sss{P}}\nonumber.
 \eea
The ghost number determines also the labeling and the indices of the
 superfields according to,
\begin{center}
\begin{tabular}{|c|c|c|c|c|c|c|c|c|c|c|c|c|c|}\hline
  superfield $K^*$& M & N & O & P & Q & R & S & T & U & V & X & Y & Z\\
  $\gh$ & 9 & 8 & 7 & 6 & 5 & 4 & 3 & 2 & 1 & 0 & -1 & -2 & -3\\
  index & Z & X & Y & A & B & C & D & E & F & G & H & I & J \\ \hline
\end{tabular}
\end{center}
The major advantage of this labeling of the superfields is that
one can directly read off what kind of superfields that should be
multiplied with any of the coefficients, just by looking at their
label and index structure. E.g. a ${\om^{\sss
  C_1}}_{\sss D_1}$-term will always multiply the superfields as $R^*_{\sss C_1}{\om^{\sss
  C_1}}_{\sss D_1}S^{\sss D_1}$ in any dimension. Furthermore, one is not confused
when interpreting expressions in different dimensions, since an
associated superfield $K^*$ of a specific ghost number will have
the same label in all dimensions - for example, the associated
superfield $S^*_{\sss D}$ will always denote a ghost number $3$
field,
regardless of the dimension of the supermanifold it lives on. 
All objects are ordered with the associated superfields to the
left and the superfields to the right.
%
%
%
%
\section{Canonical generator for superfields}\label{proof}
\begin{theorem}{
Given a canonical generator $\Ga$ of the form
  \be
    \Ga=K^*_{\sss P_1}K^*_{\sss P_2}\ldots K^*_{\sss P_n}{\Ga^{\sss
    P_1P_2\ldots P_n}}_{\sss P_{1'} \sss P_{2'}\ldots P_{n'}}K^{\sss P_{1'}}K^{\sss P_{2'}}\ldots K^{\sss
    P_{n'}},
 \ee
where $\Ga$ does not contain both a specific superfield and its
associated superfield simultaneously. This will generate a
canonical transformation of the superfields and the associated
superfields which is first order in the real and even parameter
$\ga$ that parametrizes the transformation.
 }
\end{theorem}
\begin{proof}
Consider arbitrary superfields $K^{\sss{P}}$ and associated
superfields $K^*_{\sss{P}}$.
Let $\bar{K}^{\sss{P}}$ and $\bar{K}^*_{\sss{P}}$ denote the
corresponding canonically transformed superfields. Since $\Ga$ does not
contain both $K^{\sss P}$ and $K^*_{\sss P}$ simultaneously, we have
$(\Ga,\Ga)_{\sss n}=0$. This implies that $\Ga$ contracted with $K^{\sss P}$
and $K^*_{\sss P}$ must satisfy $((K^{\sss P},\Ga)_{\sss n},\Ga)_{\sss n}=0$
and $((K^*_{\sss P},\Ga)_{\sss n},\Ga)_{\sss n}=0$. Thus, either
we have the superfields transforming as $\bar{K}^{\sss{P}} = K^{\sss
  P}\,\,\mbox{and}\,\,\bar{K}^*_{\sss{P}}= e^{\ga \mbox{\small
    ad}\,\Ga}K^*_{\sss P} = K^*_{\sss{P}}+\ga(K^*_{\sss{P}},\Ga)_{\sss
  n}$ (since higher order terms vanish) or $\bar{K}^{\sss{P}} = e^{\ga
  \mbox{\small ad}\,\Ga}K^{\sss P}= K^{\sss{P}}+\ga(K^{\sss{P}},\Ga)_{\sss n}\,\,\mbox{and}\,\,
\bar{K}^*_{\sss{P}}=K^*_{\sss{P}}$, or they stay untransformed
depending on whether $\Ga$ contains the superfields or not. Notice
that the $n$-bracket is preserved by the canonical transformations
generated by $\Ga$,
$(\bar{K}^{\sss{P}},\bar{K}^*_{\sss{P'}})_{\sss
n}={\de^{\sss{P}}}_{{\sss{P'}}}$.
\end{proof}
It follows as a trivial corollary that for every local action $S_0$
with a finite number of polynomial terms, $S_{\Ga}=e^{\textrm{\small
ad}\Ga}S_0$ will also consist of a finite number of terms.
%
%
%
%
\section{Solution to the six dimensional model}\label{container}
Below is given the full solution (with arbitrary parities of the
fields) to the master equation for the $n=6$ models studied in
section \ref{sixdimensional}. Here we denote Grassmann parities as
$\ep_{\sss{C}}=\scriptstyle{C}$,
 \bea
 {\om^{\sss{C_1}}}_{\sss{C_2} \sss{C_3}}&=&\al {\om^{\sss{C_1}}}_{\sss{D}}{\ga^{\sss{D}}}_{\sss{C_2} \sss{C_3}}\nn,\\
 {\om^{\sss{D_1}}}_{\sss{C_1} \sss{C_2} \sss{C_3}}&=&-2\al^2{\ga^{\sss{D_1}}}_{\sss{C_1} \sss{C}}{\om^{\sss{C}}}_{\sss{D}}{\ga^{\sss{D}}}_{\sss{C_2}
   \sss{C_3}}(-1)^{\sss{C_1}(\sss{C+}\sss{C_2+}\sss{C_3+}\sss{1})}\nn,\\
 {\om^{\sss{D_1}}}_{\sss{C_1} \sss{D_2}}&=&-2\al{\ga^{\sss{D_1}}}_{\sss{C_1}
   \sss{C}}{\om^{\sss{C}}}_{\sss{D_2}}(-1)^{\sss{C_1}\sss{(C+}\sss{D_2}+\sss{1)}}\nn\\
 &&{}-4\ga\om^{\sss{D_1} \sss{D}}\ga{\sss{D} \sss{D_2}
   \sss{C_1}}(-1)^{\sss{D_1}(\sss{D+1)+}\sss{C_1}\sss({D_2+D})\sss{+D_2 D}}\nn,\\
 \om_{\sss{D_1} \sss{D_2} \sss{D_3}}&=& -\ga \ga{\sss{D_1} \sss{D_2}
   C}{\om^{\sss{C}}}_{\sss{D_3}}(-1)^{\sss{(D_1}\sss{+D_2)}\sss{(C+D_3+1)}}\nn,\\
 \om_{\sss{D_1} \sss{D_2} \sss{C_1}\sss{C_2}}&=&{}-\ga\al\ga{\sss{D_1} \sss{D_2}
  \sss{C}}{\om^{\sss{C}}}_{\sss{D}}{\ga^{\sss{D}}}_{\sss{C_1}\sss{C_2}}(-1)^{\sss{(D_1+}\sss{D_2)(1+C+}\sss{C_1+}\sss{C_2)}}\nn\\
&&{}-4\ga^2{\ga}_{\sss{C_1} \sss{D_1 D}}\om^{\sss{D
D'}}{\ga}_{\sss{C_2 D_2}\sss{D'}}\times\nn\\
&&{}\times(-1)^{\sss{D(D'+1)+(D'+C_2+D_2)(C_1+D_1)+C_1D_1+D_2(C_1+C_2)}}\nn\\
  &&{}+4\al\ga\ga_{\sss{C_1} \sss{D_1 D}}{\ga^{\sss{D}}}_{\sss{C_2
  C}}{\om^{\sss{C}}}_{\sss{D_2}}\times\nn\\
&&{}\times(-1)^{\sss{C_1(D_1+D_2)+}\sss{(C_1+D_1)(D+}\sss{C_2+}\sss{D_2+1)+}\sss{C_2(C+1)}}\nn,\\
  \om_{\sss{C_1} \sss{C_2} \sss{C_3}\sss{C_4}\sss{C_5} \sss{C_6}}&=&-5\al\bet\ga_{\sss{C_1} \sss{C_2} \sss{C_3} \sss{C_4}
    \sss{C}}{\om^{\sss{C}}}_{\sss{D}}{\ga^{\sss{D}}}_{\sss{C_5} \sss{C_6}}(-1)^{\sss{(C_1+}\sss{C_2+}\sss{C_3
    +}\sss{C_4)(C+1+}\sss{C_5+}\sss{C_6)}}\nn,\\
  \om_{\sss{C_1} \sss{C_2} \sss{C_3} \sss{C_4}
  \sss{D_1}}&=&-5\bet\ga_{\sss{C_1} \sss{C_2} \sss{C_3} \sss{C_4
  C}}{\om^{\sss{C}}}_{\sss{D_1}}(-1)^{\sss{(C+}\sss{D_1+1)(}\sss{C_1+}\sss{C_2+}\sss{C_3+}\sss{C_4)}}\nn\\
  &&{}+4\ga\al^2\ga_{\sss{C_1} \sss{D_1 D}}{\ga^{\sss{D}}}_{\sss{C_2
    C}}{\om^{\sss{C}}}_{\sss{D'}}{\ga^{\sss{D'}}}_{\sss{C_3}
    \sss{C_4}}\times\nn\\
& &{}\times(-1)^{\sss{(C_1+}\sss{D_1)(D+}\sss{C_2+1+}\sss{C_3+}\sss{C_4)+}\sss{C_2(C+}\sss{C_3+}\sss{C_4+1)}}\nn.\\
 \eea
\end{appendix}
\bibliographystyle{utphysmod2}
\bibliography{biblio1}
\end{document}